\documentclass[fleqn,usenatbib]{mnras}

\usepackage{newtxtext,newtxmath}

\usepackage[T1]{fontenc}

\DeclareRobustCommand{\VAN}[3]{#2}
\let\VANthebibliography\thebibliography
\def\thebibliography{\DeclareRobustCommand{\VAN}[3]{##3}\VANthebibliography}

\usepackage{graphicx}	
\usepackage{amsmath}	
\usepackage{dsfont}
\usepackage{orcidlink}
\usepackage{bm}
\usepackage{balance}

\title[Flux scale errors in synchrotron maps]{Bayesian evidence for flux scale errors in Galactic synchrotron maps}

\author[M. J. Wilensky et al.]{
Michael J. Wilensky\,\orcidlink{0000-0001-7716-9312},$^{1, 2, \dag, \parallel}$\thanks{E-mail: michael.wilensky@mcgill.ca}
Melis O. Irfan\,\orcidlink{0000-0003-2021-7357},$^{3,4}$
Philip Bull\,\orcidlink{0000-0001-5668-3101},$^{2,4}$
\\
$^{1}$Department of Physics and Trottier Space Institute, McGill University, 3600 University Street, Montreal, QC H3A 2T8, Canada\\
$^{2}$Jodrell Bank Centre for Astrophysics, University of Manchester, Manchester, M13 9PL, United Kingdom\\
$^{3}$Institute of Astronomy, Madingley Road, Cambridge CB3 0HA, UK\\
$^{4}$Department of Physics and Astronomy, University of Western Cape, Cape Town 7535, South Africa\\
$^{\dag}$CITA National Fellow\\
$^{\parallel}$TSI Postdoctoral Fellow
}

\pubyear{\the\year{}}

\begin{document}
\label{firstpage}
\pagerange{\pageref{firstpage}--\pageref{lastpage}}
\maketitle

\begin{abstract}
The 408 MHz Haslam map is widely used as a low-frequency anchor for the intensity and morphology of Galactic synchrotron emission. Multi-frequency, multi-experiment fits show evidence of spatial variation and curvature in the synchrotron frequency spectrum, but there are also poorly-understood multiplicative flux scale disagreements between experiments. We perform a Bayesian model comparison across a range of scenarios, using fits that include recent spectroscopic observations at $\sim 1$~GHz by MeerKAT as well as a reference map from the OVRO-LWA at 73 MHz. In the few square degrees that we analyzed, a large uncorrected flux scale factor potentially as large as 1.6 in the Haslam data is preferred, indicating a 60\% overestimation of the brightness. This partly undermines its use as a reference map. We also find that models with nonzero spectral curvature are statistically disfavored. Given the limited sky coverage here, we suggest a similar analysis across many more regions of the sky to determine the extent and variation of flux scale errors, and whether they should be treated as random or systematic errors in analyses that use the Haslam map as a template. 
\end{abstract}




\section{Introduction}

Synchrotron emission from our own Galaxy is a pernicious contaminant of cosmological surveys at radio and microwave wavelengths, particularly those making total intensity (and polarisation) maps of the Cosmic Microwave Background and 21cm line emission \citep{Tegmark2000,planck15x, Planck2020, Liu2020}. Fortunately, Galactic synchrotron emission is expected to derive from free electron populations with power-law energy distributions, leading to a spectral energy distribution (SED) along each line of sight that closely approximates a power-law in frequency, $S_\nu \propto (\nu / \nu_{\rm ref})^\alpha$ \citep{RAText}. Observed values of the power-law spectral index tend to reside in the range $-1.2 \lesssim \alpha \lesssim -0.5$, with some variation as a function of Galactic latitude \citep{planck15xxv}. This permits a simple model of synchrotron foreground emission to be constructed and subtracted from radio and microwave data, revealing the cosmological maps of interest, or other statistics like the power spectrum, up to some corrections for signal loss and residual contamination depending on the foreground removal method that is used \citep{Planck15ix, planck15x, Cheng2018, Mertens2018, Sims2019, Planck2020, Cunnington2021, Kern2021}.

An important and long-standing ingredient of many foreground removal approaches is the Haslam 408~MHz all-sky survey \citep{Haslam82}. The Haslam map anchors the overall amplitude of the synchrotron power-law model in each direction on the sky. Assuming a power-law SED and a suitable model of the spectral index variation across the sky, the synchrotron intensity can then be predicted across a wide range of frequencies. Multiple sky models, such as the Planck Sky Model \citep{psm} and Python Sky Model \citep{pysm}, rely on the Haslam data as a proxy for all-sky synchrotron emission amplitude in this way. Additionally, emission models such as the Global Sky Model \citep{gsm}, which use empirical data across a wide range of frequencies, still rely on the Haslam data to provide a low frequency, high angular resolution (56 arcmin) synchrotron emission estimate. Similarly, Bayesian fits to empirical data, such as the \texttt{Commander} fits to the \emph{WMAP} and \emph{Planck} data \citep{planck15x}, use the Haslam data as a prior for synchrotron emission amplitude, with a normalization factor that changes this amplitude across their frequency range. SEDs constructed from multiple experiments across frequencies from tens of MHz to tens of GHz have detected a small degree of curvature in the power-law along some lines of sight \citep{plat98, costa08, kogut12, edges19, Irfan2022}, and a frequency-dependent correction to the spectral index of the form $\alpha \to \alpha + C \log(\nu / \nu_{\rm ref})$ has been suggested by \citet{kogut12}. This hints at energy losses through inverse Compton scattering for the relativistic charged particles responsible for synchrotron emission within the Galactic magnetic field \citep{Strong11}, and also has implications for the accuracy of the synchrotron model at frequencies significantly higher and lower than 408~MHz.

There are known flaws in the Haslam map however. For example, the original map contains unsubtracted (or imperfectly subtracted) point sources, and striping caused by imperfect filtering of correlated noise in the time-ordered data. Many of these issues were addressed in a reprocessing of the original data \citep{Remazeilles:2014mba}, but some hints of residual systematic effects remain. For instance, \citet{planck15x} note that the synchrotron emission amplitude at 408\,MHz requires  correcting for use at higher frequencies if the spectral index is to be treated as constant over those frequencies, while \citet{planck15xxv} note that a single spectral index value across all frequency ranges cannot be used to fit empirical data without generating ``large and spurious [flux scale] corrections at 408 MHz.'' However, \citet{Remazeilles:2014mba} point out that there is a large possible source of calibration uncertainty associated with the Haslam data in the form of an unknown relationship between the angular scale to which the map was calibrated and the brightness temperature scale, due to power outside of the main lobe of the primary beam. This beam efficiency factor can be upwards of 30\% for a radio experiment \citep{PlanckBeamFactor, Du2016}. An important question is whether such issues could have led to incorrect inferences in some of the (many) studies that use the Haslam map as an anchor, e.g. regarding the presence of curvature in the synchrotron SED, or over-/under-subtraction of foreground contamination.

In this paper, we use the tools of Bayesian model comparison and new {\it spectroscopic} radio observations at $\sim 1$~GHz to assess how errors in the overall flux scale may restrict the ability of sub-GHz data alone to constrain the synchrotron spectral index curvature. We apply an analysis method similar in spirit to a Bayesian jackknife framework called \texttt{Chiborg} presented by \citet{Chiborg}, wherein an analyst proposes a discrete selection of hypotheses to be compared in a ``switchboard" fashion.  Formally, this switchboard allows the analyst to turn features in the model on or off by adjusting hyperparameters in a Bayesian hierarchical model \citep{Gelman2021}. The marginal likelihood, often called the \textit{evidence}, is then calculated for each hypothesis, and these can be normalized to create a posterior probability mass function over them if one is willing to assign prior probabilities to the various hypotheses. This, in turn, can be used to select the set of hypotheses that best explains the data. 

We restrict ourselves to considering a flux scale error for each survey dataset used in our analysis as the only systematic effect in the experiments, as hypothesized by \citet{planck15xxv}. In other words, we constrain the spectral parameters of the underlying Synchrotron emission while allowing for the possibility that experiments' reported brightness maps may need to be multiplicatively rescaled in order to mutually agree. The values of these flux scale factors are tempered by a prior based on each experiment's reported uncertainty regarding the calibration of their flux scale. We then also test whether priors reflecting larger flux scale uncertainty are supported by the data in terms of marginal likelihood. Since the underlying model for the data prohibits analytic marginalization, we use the nested sampler \texttt{polychord} \citep{PolychordI, PolychordII} to simultaneously calculate the evidence for each hypothesis and constrain free parameters such as the size of the flux scale factors.

While it has long been known that the Haslam map may have an error in its overall flux scale that exceeds the quoted 10\% uncertainty \citep[e.g.][]{Remazeilles:2014mba, Monsalve2021}, in this paper we show that there is strong statistical evidence for an overall flux scale error that may be as high as ~60\% in the small region we analyze. Careful reading of \citet{Haslam1981} suggests that the flux scale error may exhibit some spatial variation across the map, and so this figure may not reflect the map as a whole. However, an error of this size has potentially serious implications for analyses that rely on the Haslam map, such as joint CMB foreground component separation \citep{PlanckBeamFactor, planck15x, planck15xxv, BeyondPlanck1}, and foreground removal strategies in 21-cm experiments \citep{Eastwood2019, Pagano2024}. 

In \S\ref{sec:methods} we formally describe the data, models, statistical methods, and analysis framework. In \S\ref{sec:hyp_comp}, we present the model selection results, while \S\ref{sec:fits} shows the resulting posterior inferences within several of the models considered. We also compare the inferences to our physical understanding of the situation and previous analyses that asked similar questions. We conclude in \S\ref{sec:conc}.

\section{Methods}

Here we discuss the mathematical model of the data, including spectral behavior, thermal noise, and flux scale uncertainties. We also present an array of priors, which reflect different underlying hypotheses (i.e. assumptions) about the experimental and astrophysical circumstances.

\label{sec:methods}

\subsection{Synchrotron SED data and models}
\label{sec:data}

For this analysis we use the same frequency range and regions on the sky as probed by \citet{Irfan2022}: the OVRO Long Wavelength Array (OVRO-LWA) 73\,MHz data, the Haslam 408\,MHz data, and the MeerKLASS 971 -- 1075\,MHz data within five regions of radius $1.8^\circ$ centered at (RA, Dec) = (161$^\circ$, 2.4$^\circ$), (164$^\circ$, 2.7$^\circ$), (167$^\circ$, 3.5$^\circ$), (157.2$^\circ$, 3.6$^{\circ}$), and (155.7$^\circ$, 2.1$^\circ$). In Galactic coordinates, ($l$, $b$), these are (246.5$^{\circ}$, 50.7$^{\circ}$), (249.4$^{\circ}$, 53.1$^{\circ}$), (252.1$^{\circ}$, 55.8$^{\circ}$), (241.3$^\circ$, 48.6$^\circ$), and (241.6$^\circ$, 46.5$^\circ$). We denote the first three \textit{primary} regions as fields 0, 1, and 2. We denote the last two \textit{reference} regions as fields I and II. The reference fields were chosen since they are relatively dim regions in all three experiments' maps, and because a choice of dim region makes the analysis most similar to that performed by \citet{Irfan2022}. 

The data have all been smoothed to a common resolution of $1.8^\circ$ with Gaussian beams. Both the OVRO-LWA \citep{Eastwood2018} and MeerKLASS \citep{meerklass} experiments are recent (within the last decade) 21cm intensity mapping endeavors; the former is interferometric, while the latter is used in single-dish mode. For the Haslam values, we use the destriped but \textit{not} desourced map from \citet{Remazeilles:2014mba}, since neither the MeerKLASS nor OVRO-LWA measurements have been desourced \citep{Irfan2022}. We use an aperture photometry method to extract SED difference data points at each frequency from the available maps (explained below). This is an important step since there are modes on large spatial scales that are missing in the OVRO-LWA and MeerKAT maps\footnote{resulting in negative surface brightness in the maps} that are present in the Haslam map.

\begin{figure}
    \centering
    \includegraphics[width=\linewidth]{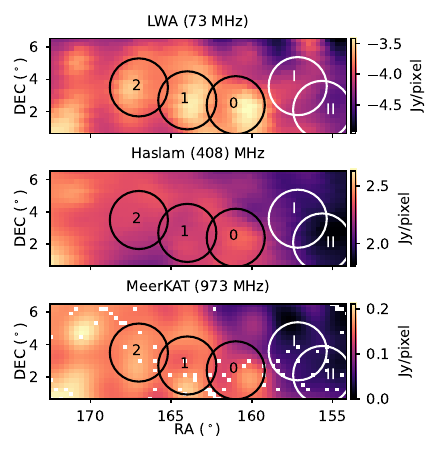}
    \caption{Smoothed maps from each experiment with primary apertures (black circles) and reference apertures (white circles) drawn. Each aperture has a diameter of $1.8^\circ$. Each pixel is $0.3^\circ$ on a side. White pixels in the MeerKAT map represent data lost to RFI flagging \citep{Irfan2022}. The units are in terms of the flux density per $0.3^\circ\times 0.3^\circ$ pixel, as calculated by Equation 13 of \citet{Irfan2022} using the Rayleigh-Jeans law. Due to missing large scale modes, the OVRO-LWA map has regions of negative surface brightness in this field. The MeerKAT map is also missing large-scale modes due to a mean-centering procedure to remove temperature contributions from the ground spillover and receiver temperature.  All pixels in the MeerKAT map also have a constant offset by the dimmest pixel applied \citep{Irfan2022}. As explained in the main text, our aperture photometry procedure is not sensitive to constant offsets.}
    \label{fig:fields}
\end{figure}

The apertures are shown in Figure \ref{fig:fields}. In typical applications of aperture photometry, the aperture is chosen to relatively tightly surround a compact source, and an annulus is drawn around the source so as to enclose a representative sample of the larger-scale background (or foreground) that is not specifically associated with the source of interest. The average (or e.g. median) over this annulus is then subtracted from the brightness in the aperture to isolate the contribution of the source of interest \citep[e.g.][]{PlanckBeamFactor}. In our case, the source of interest is diffuse Galactic synchrotron emission, and there are no clear annuli that we can draw around our apertures to determine the contribution of the modes on large spatial scales independent of the modes on small ones. Instead, we form two data sets consisting of three spectrum differences each by subtracting the mean surface brightness in a reference aperture from the mean surface brightness in the primary apertures at every frequency. We use these two data sets to perform two parallel analyses, corresponding to each choice of reference field, in order to examine the robustness of our conclusions. We explain the nature of the parallel analyses in more detail at the end of \S\ref{sec:model_comp_framework} once the mathematical models in consideration have been defined.

In the usual annulus-based method, one assumes that what is left over is just the SED of the compact source, since the background estimate is chosen locally enough to have effectively been subtracted. In our case, supposing any given aperture is dominated by Galactic synchrotron emission, we suggest that the difference between any two apertures is modeled by a difference between two (potentially curved) power laws. In other words, the spectrum in the $i$th aperture is modeled as
\begin{equation}
    S_i(\nu) = S_i(\nu_0)\left(\frac{\nu}{\nu_0}\right)^{\alpha_i(\nu_0) + c_i \ln \left(\nu/\nu_0\right)} + \text{ large scale contributions}
    \label{eq:SED}
\end{equation}
where $\nu_0$ is a reference frequency, $S_i(\nu_0)$ is the surface brightness at the reference frequency, $\alpha_i(\nu_0)$ is the spectral index at that frequency, and $c_i$ is the curvature of the power law. We have experimented with different choices of reference frequency and have found that the choice is mostly inconsequential; the choice of 408 MHz allows one to easily compare inferred values of $S_i(\nu_0)$ to the original maps, however. Since all the fields are fairly close (quantified below, see Figure \ref{fig:fields}), a difference between two fields should subtract out the modes on large spatial scales. We therefore make use of an SED difference, $\Delta m_{ij}(\nu)$, which we define as the difference between two SEDs when contributions from large spatial scales are ignored:
\begin{equation}
    \Delta m_{ij}(\nu) \equiv S_i(\nu_0)\left(\frac{\nu}{\nu_0}\right)^{\alpha_i(\nu_0) + c_i \ln \left(\nu/\nu_0\right)} - S_j(\nu_0)\left(\frac{\nu}{\nu_0}\right)^{\alpha_j(\nu_0) + c_j \ln \left(\nu/\nu_0\right)}.
    \label{eq:sed_diff}
\end{equation}
In other words, we are assuming the maps have complete information on spatial scales between $1.8^\circ$ and roughly $12^\circ$ (roughly the maximum distance between analysis and reference apertures), such that the differencing procedure on the smoothed maps is probing the same physical processes. If the sampling of these spatial scales is incomplete i.e. if the map is missing power on those spatial scales, then this may cause errors in the difference relative to other experiments even in the absence of a flux scale error (e.g. smaller differences between fields than a power law frequency scaling of both fields would predict). These errors might then manifest as an observed anomaly in our analysis. 

Since the Haslam map was made from autocorrelation surveys and has a calibrated zero-level, it probably has complete sampling of the Galactic synchrotron emission on large spatial scales. We are more uncertain about the MeerKAT and OVRO-LWA maps. The MeerKAT maps do not have calibrated zero levels and were mean-centered by \citet{Irfan2022} in order to avoid modeling the ground spillover and atmospheric noise. This mean was over a larger spatial scale than what we rely on in this analysis. The OVRO-LWA map was made using a minimum spherical harmonic $\ell_\mathrm{min} = 1$, whereas the $\sim 12^\circ$ separation in Right Ascension between our apertures corresponds to Nyquist sampling the $m=15$ azimuthal modes at nearly constant Declination. Modes with azimuthal mode number less than $m=15$ may still have significant variation across 12$^\circ$ of Right Ascension, but if the choice of $\ell_\mathrm{min}=1$ was meant to suggest that sampling is reliable down to that mode, then the spatial scale of the differencing procedure should not be an issue. On the other hand, comparisons within \citet{Eastwood2018} to maps from the LWA1 Low Frequency Sky Survey \citep[LLFSS;][]{Dowell2017} demonstrate $10-20\%$ disagreements between the two maps\footnote{Here we focus on the comparison at 70 MHz within \citet{Eastwood2018}. The discrepancies vary slightly in strength and morphology as a function of frequency.} with visible features on spatial scales of roughly a few degrees. This could be a sign of missing flux in one or both of the OVRO-LWA and LLFSS maps. We therefore list the potential for incomplete spatial sampling to present an observed anomaly as an important caveat of this work. Nevertheless, the isotropic background at each frequency, which by definition contributes identically to all of the apertures but whose level and frequency scaling is currently debated \citep{background}, should be successfully subtracted out by this differencing procedure. 

\subsection{Flux scale and noise uncertainties}

Each experiment quotes a fractional uncertainty regarding the calibration of their flux scale. OVRO-LWA quotes between a 5\% and 10\% flux scale uncertainty based on the flux density measurement of 11 known calibrators, including Cygnus A, Cassiopeia A, and Virgo A. The Haslam map flux scale uncertainty is often quoted as 10\%, which relates to the level of scatter seen when correlating this data set with an absolutely calibrated map at 404 MHz \citep{map404}. MeerKLASS calibrate their data with respect to a system temperature model which consists of Galactic and terrestrial contributions; if this calibration were perfect, the difference between the data and the model would simply be Gaussian distributed noise centered around zero. As they observe data-model residuals of 0.3K for a 16K system temperature model, they quote their calibration uncertainty as good to 2\%. The flux scale uncertainties we choose to carry forward for OVRO-LWA, Haslam map, and MeerKLASS are 5\%, 10\% and 2\%, respectively. 

All of the calibration descriptions in the original works producing these data suggest that there may be some spatial dependence in the overall flux scale errors, but it is unclear what the spatial scale of the variations is. This is to say, the authors report that their maps appear too bright in some regions and too dim in others, but in a way that is inconsistent with their estimate of variations from thermal noise (which would vary on spatial scales at the map's resolution). However, in a preliminary analysis when we compared independent flux scale errors per field versus a single rescaling for all the fields, we found the single rescaling model to be vastly preferred statistically. Based on each experiments' flux scale calibration description, we expect that portions of the map within a few degrees of each other would have similar flux scale errors. We therefore present an analysis where we assume the measured SED difference with respect to primary aperture $i\in\{0, 1, 2\}$ and reference aperture $j\in\{\mathrm{I},\mathrm{II}\}$ are given by
\begin{equation}
    d_{ij}(\nu) = g(\nu) \Delta m_{ij}(\nu) + n_i(\nu) - n_j(\nu),
    \label{eq:model_spec}
\end{equation}
where $n_i(\nu)$ is a thermal noise fluctuation in aperture $i$, and $g(\nu)$ is a single frequency-dependent \textit{flux scale factor} that applies equally to all the fields (i.e. to the entire $18.3^\circ \times 6.3^\circ$ map that we analyze).\footnote{While technically we analyze quantities that are surface brightnesses rather than flux densities, the effect of dimensionless multiplicative factors on both quantities is the same, so we retain the ``flux scale'' terminology throughout.} This dimensionless parameter represents the overall flux scale calibration of each experiment. If this factor is equal to 1, it means that the underlying SED difference does not need to be rescaled in order to fit the experimental data. If it is significantly different from 1, it indicates that there is a \textit{flux scale error} in the sense that the reported brightness does not match the underlying physical emission. The frequency dependence encodes which experiment the factor belongs to. 

We do not model independent flux scale factors for each of the 148 MeerKAT frequency channels. Instead, motivated by an apparent discontinuity in the spectra in Figure 11 of \citet{Irfan2022}, we break the full band into two subbands and fit two independent factors, one for each subband. The division point was chosen by eye at the 60th frequency channel, with center frequency 1007.34 MHz, which is roughly 2/5 the way through the full MeerKLASS band. This discontinuity is part of a known bandpass feature with a standard correction strategy discussed by \citet{Wang2021}, which we do not apply for reasons discussed in \S\ref{sec:wide_haslam}. The flux scale factor for a subband rescales all of the MeerKAT data in that subband. 

This makes for four total flux scale factors, which we denote $g_\text{LWA}$, $g_\text{Has}$, $g_\text{MK1}$, and $g_\text{MK2}$, and bundle into the vector $\mathbf{g}$. To map from these flux scale factors back to a more manifestly frequency-dependent form, one may think of the data (and model) in the discrete frequency channels as elements of a (block-)vector space. To illustrate this, we define the block matrix
\begin{equation}
    \mathbf{A} \equiv 
    \begin{pmatrix}
    1 &0 &0 &0 \\
    0 &1 &0 &0 \\
    \mathbf{0}_n &\mathbf{0}_n &\mathbf{1}_n &\mathbf{0}_n \\
    \mathbf{0}_m &\mathbf{0}_m &\mathbf{0}_m &\mathbf{1}_m 
    \end{pmatrix}
\end{equation}
where $\mathbf{0}_n$ is a column vector of $n$ zeros (the length of the first MeerKAT subband), and similarly for $\mathbf{1}_n$, etc. We also define
\begin{equation}
    \mathbf{B} \equiv 
    \begin{pmatrix}
        \mathbf{I}_3 \otimes \mathbf{I}_{N_\nu} & -\mathbf{1}_3 \otimes \mathbf{I}_{N_\nu}
    \end{pmatrix}
\end{equation}
where $\mathbf{I}_k$ is the identity matrix in $k$ dimensions, $N_\nu = n + m + 2$ is the total number of frequencies, and $\otimes$ denotes the Kronecker product. Finally, defining $\Delta\mathbf{M}_{ij}$ as a diagonal matrix whose $k$th diagonal entry is equal to $\Delta m_{ij}(\nu_k)$, and further defining
\begin{equation}
    \Delta\mathbf{M}_j \equiv 
    \begin{pmatrix}
        \Delta\mathbf{M}_{0j} \\
        \Delta\mathbf{M}_{1j} \\
        \Delta\mathbf{M}_{2j}
    \end{pmatrix},
    \label{eq:Mj}
\end{equation}
as well as
\begin{equation}
    \tilde{\mathbf{n}}_j \equiv
    \begin{pmatrix}
        \mathbf{n}_0 \\
        \mathbf{n}_1 \\
        \mathbf{n}_2 \\
        \mathbf{n}_j
    \end{pmatrix},
\end{equation}
we can write the data vector for when reference field $j$ is used as
\begin{equation}
    \mathbf{d}_j \equiv \Delta\mathbf{M}_j \mathbf{A} \mathbf{g} + \mathbf{B}\tilde{\mathbf{n}}_j
    \label{eq:diff_vec}
\end{equation}

We develop a (Gaussian) noise covariance model for the OVRO-LWA and Haslam maps by propagating uncorrelated (i.e. diagonal) pixel-pixel covariance matrices at our pixel resolution ($0.3^\circ$) through a linear smoothing function that emulates the mapping and subsequent smoothing process, broken into three steps. First, we propagate the uncorrelated covariance to each experiments' reported resolution, so that it has a correlation structure within our pixelization scheme that matches the experiments' map resolution. The smoothing kernel for each experiment is Gaussian with full-width-half-max equal to the reported resolution of the corresponding experiment. At this point, the matrices need to be rescaled so that they have the thermal variance that the experiments report at this resolution. Finally, we smooth once more to the $1.8^\circ$ resolution of the analysis to emulate our pre-processing. This last step is the same smoothing function we use to smooth the maps to the common resolution. For the MeerKAT map, we essentially do the same but starting with a map of the system temperature. We can then formally propagate these covariance matrices through the aperture photometry procedure. This allows us to assess the relative correlation from thermal noise within the fields in the SED differences. In other words, it allows us to analytically calculate, for arbitrary fields $i$ and $j$, 
\begin{equation}
    \mathbf{N}_{ij} \equiv \langle\mathbf{n}_i\mathbf{n}_j^T\rangle
    \label{eq:N_ij}
\end{equation}
where $\mathbf{n}_j^T$ is the transpose of the vector $\mathbf{n}_j$ and $\langle\cdot\rangle$ indicates an ensemble expectation over possible noise realizations.

This model predicts that thermal noise in neighboring fields is lightly correlated (correlation coefficient less than 0.2). Since all of the primary fields are independent of the reference field, but a common reference is used, this ends up contributing no more than 0.1 to the total correlation coefficient between any two fields for any given experiment. For simplicity, we ignore these correlations between neighboring fields, but include the induced correlation from using a common reference field (which produces a correlation coefficient of $\sim 1/2$ across the 3 primary fields). Bundling the SED parameters as $\boldsymbol{\theta}$ and treating $\tilde{\mathbf{n}}_j$ as a Gaussian random vector in Equation \ref{eq:diff_vec} gives the likelihood function in the form
\begin{equation}
    \mathbf{d}_j | \boldsymbol{\theta}, \mathbf{g}, \tilde{\mathbf{N}}_j \sim \mathcal{N}\left(\Delta\mathbf{M}_j\left(\boldsymbol{\theta}\right) \mathbf{A} \mathbf{g}, \mathbf{B}\tilde{\mathbf{N}}_j\mathbf{B}^{T}\right),
    \label{eq:like_def}
\end{equation}
where $\mathcal{N}(\mathbf{\boldsymbol{\mu}}, \mathbf{\Sigma})$  represents a multivariate Gaussian distribution with mean vector, $\boldsymbol{\mu}$, and covariance matrix, $\mathbf{\Sigma}$, $\tilde{\mathbf{N}}_j = \langle\tilde{\mathbf{n}}_j\tilde{\mathbf{n}}_j^T\rangle$. Unpacking $\tilde{\mathbf{N}}_j$ more explicitly, one may use properties of the Kronecker product to show
\begin{equation}
    \mathbf{B}\tilde{\mathbf{N}}_j\mathbf{B}^{T} = 
    \mathbf{N}_p
     + (\mathbf{1}_3\mathbf{1}_3^T) \otimes \mathbf{N}_{jj}
     \label{eq:final_cov}
\end{equation}
where $\mathbf{N}_p$ is the joint covariance of the primary fields (cf. Equation \ref{eq:N_ij}):
\begin{equation}
    \mathbf{N}_p \equiv \begin{pmatrix}
        \mathbf{N}_{00} & \mathbf{N}_{01} & \mathbf{N}_{02} \\ \mathbf{N}_{10} & \mathbf{N}_{11} & \mathbf{N}_{12}  \\
        \mathbf{N}_{20} & \mathbf{N}_{21} & \mathbf{N}_{22} 
    \end{pmatrix}.
    \label{eq:N_p}
\end{equation}
The second term in Equation \ref{eq:final_cov} adds $\mathbf{N}_{jj}$ to every $N_\nu \times N_\nu$ block of the matrix in Equation \ref{eq:N_p}, resulting in the aforementioned correlation coefficient of $\sim 1/2$ if neighboring field correlations are ignored (whose effects are seen to be lessened by this very addition). 

Using nothing other than the information that the respective experiments' flux scale calibration is thought to be accurate to within a certain fractional percentage, we by default assume a Gaussian prior on the flux scale factors centered at 1, with variances equal to the square of the fractional uncertainties reported from each experiment.\footnote{One can combine this with Equation \ref{eq:diff_vec} to marginalize over the per-experiment flux scale factors and work with a low-dimensional likelihood. This allows one to incorporate the uncertainty in these parameters at the expense of being unable to explicitly infer them. By definition, the marginal distribution of the SED parameters would be the same, and one could implicitly observe the need for non-unity flux scale factors, which was essentially the motivation for this work \citep{Irfan2022}.} 
\begin{equation}
    \mathbf{g} \sim \mathcal{N}\left(\mathbf{1}_4, diag\left(\sigma^2_\text{LWA},\sigma^2_\text{Has},\sigma^2_\text{MK1},\sigma^2_\text{MK2}\right)\right)
\end{equation}
In this analysis, if we infer a flux scale factor for an experiment that deviates from 1 with statistical significance, we say that we have inferred a flux scale error. We also refer to the amount that a flux scale factor deviates from 1 as a \textit{flux scale factor offset}. We experiment with alternate flux scale factor priors (i.e. different assumptions of flux scale uncertainty) within our model comparison framework, discussed next.

\subsection{Model comparison framework}
\label{sec:model_comp_framework}

We examine a few different models using Bayesian model comparison, wherein we calculate the posterior probability of the models, $P(\mathcal{H}|\mathbf{d}_j)$, and analyze the data in light of what is most likely. This offers a formal way of comparing models that also naturally accounts for overfitting \citep{Kass1995, Jaynes2003, Mackay2003}, though it is not without complications \citep{Gelman2021}. We think of these models as formal hypotheses about the system and denote them with the variable $\mathcal{H}$. 

In each hypothesis, we change our prior about the SED parameters in the fields. We consider four models. The most basic hypothesis may be phrased colloquially as ``rather shallow and rather steep spectral indices are allowed, we have no preference about where they lie, and there is no spectral curvature.'' Thinking of this as a ``null hypothesis'', we can denote this with the priors
\begin{align}
    &\alpha_i(\nu_0) | \mathcal{H}_0 \sim \mathcal{U}\left(-1.8, 0\right) \\
    &c_i | \mathcal{H}_0 = 0
\end{align}
where $\mathcal{U}(a, b)$ notes a uniform distribution on the interval $[a,b]$, and the equality for $c_i | \mathcal{H}_0$ indicates that we fix $c_i$ at 0 for all fields in this hypothesis. We generate three more hypotheses by negating (or modifying) one or two propositions in the null hypothesis, e.g. ``only spectral indices close to the astrophysical expectation are allowed, with zero probability outside the range $[-0.9, -0.3]$, and there is no spectral curvature.'' Denoting this hypothesis $\mathcal{H}_\alpha$, we would write
\begin{align}
    &\alpha_i(\nu_0) | \mathcal{H}_\alpha \sim \mathcal{U}\left(-0.9, -0.3\right) \\
    &c_i | \mathcal{H}_\alpha = 0
\end{align}
Finally we may modify the hypothesis by allowing for negative spectral curvature, adopting a flat prior from $-0.3$ to $0$, e.g.
\begin{align}
    &\alpha_i(\nu_0) | \mathcal{H}_{\alpha c} \sim \mathcal{U}\left(-0.9, -0.3\right) \\
    &c_i | \mathcal{H}_{\alpha c} \sim \mathcal{U}\left(-0.3, 0\right)
\end{align}
We also consider models with larger flux scale uncertainties in the Haslam map by adopting a prior with a standard deviation of 20\% instead of 10\%. For example,
\begin{equation}
    g_\text{Has} | \mathcal{H}_h \sim \mathcal{N}\left(1, 0.2^2\right)
\end{equation}
Such models allow for larger inferred flux scale errors if the likelihood lends significant support away from a flux scale factor of 1, but will otherwise be penalized in the marginal likelihood calculation due to the larger prior volume. 

We initially focus on the Haslam map for clarity of demonstration. As we show below, the most apparent violation of nominal flux scale uncertainties (i.e. inferred flux scale factors whose difference from 1 greatly exceeds the standard deviation of the prior) occurs from the Haslam data, and is thus the clearest demonstration of the effects of different assumptions (different priors) on statistical inferences of Galactic synchrotron emission. In the future development of synchrotron maps, i.e. in analyses considering much more data than this, it may be important to consider the possible mischaracterization of any contributing experiments' flux scale calibration. As such, we explore the possibility of underestimated flux scale uncertainties (prior variances that are too small) in all experiments in \S\ref{sec:wide_gain}.

For all fields (primary and reference), we adopt a uniform prior on $S_i(\nu_0)$ from 1 to 4 Jy/pixel,\footnote{This refers to the flux density per $0.3^\circ\times 0.3^\circ$ pixel, as calculated by Equation 13 of \citet{Irfan2022} using the Rayleigh-Jeans law.} based on the values of the Haslam map in those fields. This gives enough breathing room for potentially strong flux scale factors but does not allow for scenarios such as a flux scale factor much greater than 2, which would be a $10\sigma$ anomaly according to the nominal flux scale uncertainties. We note that this choice of prior completely ignores the pre-differenced aperture photometry data and results in substantially larger posterior uncertainties in all the free parameters compared to an alternate choice we detail momentarily. This increase in posterior uncertainty occurs because it is difficult to fix two SED amplitudes when only their difference is being observed. We could instead parametrize $S_i(\nu_0)$ so that $g(\nu)S_i(\nu)$ is fixed to the Haslam map aperture photometry data (akin to fixing a boundary condition in a differential equation). Such a parametrization fixes the degeneracy in the inferred SED amplitudes that arises from the differencing procedure, but produces a somewhat strange model such that the \textit{scaled} SED difference always passes through the Haslam data. We have experimented with this alternate choice, and it gives similar (but sharper) inferences compared to the flat $\mathcal{U}\left(1,4\right)$ prior. We do not present this alternate choice in this work but note it as a potentially useful tool.

By Bayes' theorem,
\begin{equation}
    P(\mathcal{H}|\mathbf{d}_j) = \frac{P(\mathbf{d}_j|\mathcal{H})P(\mathcal{H})}{\sum_\mathcal{H}P(\mathbf{d}_j|\mathcal{H})P(\mathcal{H})}
    \label{eq:hyp_post}
\end{equation}
The term $P(\mathbf{d}_j|\mathcal{H})$ is the marginal likelihood, or evidence, of the hypothesis, $\mathcal{H}$. Representing the model parameters in any given hypothesis as $\boldsymbol{\theta}$, the evidence, $\mathcal{Z}$, is given by
\begin{equation}
    \mathcal{Z} \equiv P(\mathbf{d}_j|\mathcal{H}) = \int \mathop{}\!\mathrm{d}\boldsymbol{\theta} P(\mathbf{d}_j | \boldsymbol{\theta}, \mathcal{H})P(\boldsymbol{\theta} | \mathcal{H}),
    \label{eq:marg_like}
\end{equation}
where $P(\boldsymbol{\theta}|\mathcal{H})$ is the prior for the model parameters in the hypothesis $\mathcal{H}$. The term $P(\mathcal{H})$ represents the prior probability of a given hypothesis. This allows one to disfavor different hypotheses in the selection process if they represent unlikely scenarios according to a particular state of knowledge. 

We analyze the use of each reference field independently and then discuss the relative consistency of the results. That is, we perform two separate analyses by fixing $j=\mathrm{I}$ or $j=\mathrm{II}$ in Equation \ref{eq:model_spec}, and then  jointly modeling all the primary apertures ($i= 0, 1, 2$) with that choice of reference (cf. Equation \ref{eq:Mj}). We first compare the hypotheses via their marginal likelihoods, and then examine a selection of posterior probability distributions from different models to compare their inferences. In most cases we plot the results from each choice of reference field (each parallel analysis) side-by-side, with the exception of Figures \ref{fig:SED_ref3} and \ref{fig:SED_ref4}, which are each dedicated to a separate choice of reference field.

\subsection{Numerical implementation}

The nonlinearity of the model in the power law parameters leads to an analytically difficult\footnote{An alternate parametrization in terms of the logarithm of the data and parameters can be analytically tractable if additive effects like thermal noise are ignored and convenient priors are chosen. We do not do that in this work.} evidence integral (Equation \ref{eq:marg_like}). We instead use the nested sampler \texttt{polychord} \citep{PolychordI, PolychordII} to simultaneously compute the evidence of each hypothesis and constrain the remaining degrees of freedom. For the sake of reproducibility, we state our choice of \texttt{polychord} parameter settings. \texttt{polychord}'s evidence estimation is affected by the number of live points during sampling, $N_\text{live}$, which should be scaled according to the dimensionality of the problem, $N_\text{dim}$. Different hypotheses have different dimensionality. We use $N_\text{live} = 25 N_\text{dim}$. The \texttt{polychord} parameter $N_\text{repeats}$ controls how many slice sampling steps are performed to generate a new sample, which affects the correlation length of the chain. For reliable evidences, it is recommended to use $N_\text{repeats} \sim 3N_\text{dim}$. We used $N_\text{repeats} = 5N_\text{dim}$ and found the inference to be very similar between several independent chains.

\subsection{Comments on properties of the data}

The MeerKLASS data are uniquely helpful in constraining the SEDs in two regards. First, they are at a much higher frequency than the OVRO-LWA and Haslam data. Second, they consist of finely channelized measurements (i.e. they are spectroscopic). While we only fit one flux scale factor per subband, we do not average together channels within a subband; we allow each of them to be used as a separate data point, which means that we are fitting the spectral shape of the Galactic emission using many data points within a somewhat narrow band (971-1075 MHz). The application of the flux scale factors does not change the shape within each sub-band, only the overall scale (which can differ between the two sub-bands). This means both the amplitude and slope of the MeerKLASS spectrum have a lever on the SED parameters. This gives it the ability to rule out models that are locally too steep at the MeerKLASS frequencies that would otherwise be consistent if e.g. we had averaged over the entire MeerKLASS spectrum. Since steeper spectral indices imply larger amplitudes at lower frequencies, this means the MeerKLASS data have a more robust lever against overall amplitude errors (i.e. flux scale errors) in the OVRO-LWA and Haslam data than if we included a single data point at 1 GHz.

While it is entirely possible that the data from any given experiment deviates from the underlying synchrotron model due to additive effects (e.g. RFI or bright extragalactic sources\footnote{While the maps have been smoothed to a fairly coarse effective resolution ($1.8^\circ$), but rendered in $0.3^\circ$ pixels, this does not actually \textit{remove} power from extragalactic point sources. It just delocalizes it, meaning there could be significant extragalactic contributions in our apertures.}), we do not consider that possibility in this work. In other words, any evidence that we find for a non-unity flux scale factor could be an additive bias masquerading as a rescaling in our analysis because we do not give the model any other place to locate such an inconsistency. As mentioned by \cite{Irfan2022}, using the desourced Haslam map decreases its average surface brightness in the apertures by a few per cent in each field, but this is smaller than the size of flux scale factor offsets that we find below. Since desourced maps for OVRO-LWA and MeerKAT are not available, we felt it best to not attempt desourcing ourselves and instead list this point as an important caveat to this analysis. 

\section{Hypothesis Comparison}
\label{sec:hyp_comp}

\begin{figure}
    \centering
    \includegraphics[width=0.9125\linewidth]{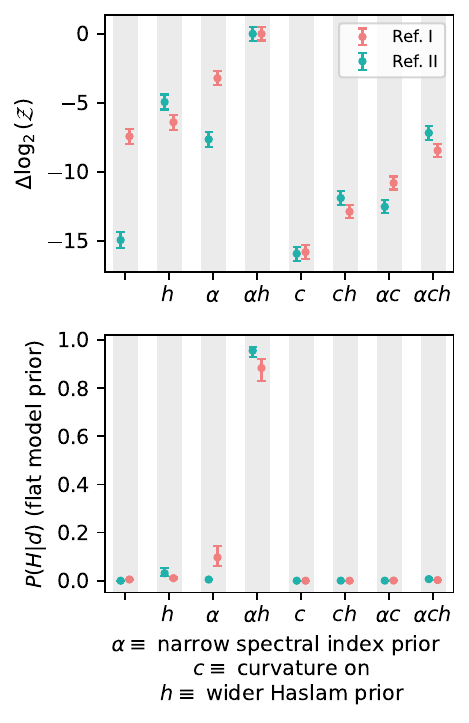}
    \caption{
    Top: the difference between the logarithm of the Bayesian evidence of the strongest hypothesis and other hypotheses, shown separately for each choice of reference field. Bottom: A posterior over hypotheses assuming a flat prior (i.e. the top on a linear scale).}
    \label{fig:hyp_post}
\end{figure}

In order to form a posterior probability over the discrete hypotheses, we must assign a prior to each hypothesis. The simplest option is to assign equal prior probability to each hypothesis, shown in Figure~\ref{fig:hyp_post}, along with $1\sigma$ Monte Carlo uncertainty estimates. This is shown on a logarithmic scale in terms of the marginal likelihoods (top panel), as well as a linear scale in terms of posterior probability (bottom panel). We use a flat prior not necessarily because we genuinely have a flat prior, but because anyone examining this plot can easily apply their own priors if we show it without a particular \textit{a priori} bias. Our choice of \texttt{polychord} settings produces roughly constant Monte Carlo uncertainties in the logarithm of the Bayesian evidence. This means the uncertainties are larger for stronger hypotheses when we put them on a linear (probability) scale as in the top panel. With a flat prior, the posterior is dominated by a single hypothesis for either reference field, which is that the Haslam map's nominal flux scale uncertainties are underestimated and steep spectral indices are disallowed ($\mathcal{H}_{\alpha h}$). All hypotheses with spectral curvature are strongly disfavored.

We quantitatively demonstrate the statistical strength of this information with a pedagogical toy model. Following the discussion of \citet{Haslam1981}, suppose that $N$ samples of the sky are compared to another survey and agreement is found with a standard deviation of 10\%. In a Normal-Inverse-Gamma model, supposing independent measurements, conditioning on a flux scale factor mean of 1, and assuming a log-flat prior on the variance for simplicity, this suggests a posterior distribution of the variance of flux scale factors of the original Haslam experiment of
\begin{equation}
    \sigma^2_\text{Has} \sim IG\left(\frac{N}{2},  \frac{N}{2} \times 0.1^2\right).
    \label{eq:IG}
\end{equation}
where $IG(a, b)$ indicates an inverse-gamma distribution with degrees of freedom equal to $a$ and scale parameter equal to $b$. This distribution can then sensibly define a prior for future analyses of the Haslam map, with the idea being that any patch of the sky has randomly sampled from the distribution of flux scale errors whose variance is known to within the uncertainty reflected by Equation \ref{eq:IG}, and whose mean is thought to be 1. Since our two hypotheses are picking out specific values of the variance, the prior odds would just be the ratio of the probability density function defined by Equation \ref{eq:IG} evaluated at the two variances. For our choice of doubling the standard deviation (quadrupling the variance), this gives a ratio of
\begin{equation}
    \log_2\left(\frac{P(\sigma^2_\text{Has}=0.1^2)}{P(\sigma^2_\text{Has}=0.2^2)}\right) = N + 2 - \frac{3N}{8}\log_2e \approx 0.46 N + 2
\end{equation}
For $N=10$, this is approximately the discrepancy between $P(\mathbf{d}_\text{II}|\mathcal{H}_{\alpha h})$ and $P(\mathbf{d}_\text{II}|\mathcal{H}_{\alpha})$ in the top panel of Figure \ref{fig:hyp_post}. The description of flux scale calibration in \citet{Haslam1981} is relatively sparse, and even in this highly oversimplified example, choosing $N$ from the given information is not straight-forward, so we are not seriously suggesting a prior of this form. However, in this analysis we have essentially sampled one realization of the flux scale error distribution, and the fact that it carries the weight of at least a few samples in that comparison suggests by itself that the error is relatively large in this section of the map. We examine the inferred flux scale factor in \S\ref{sec:fits}.

Developing a prior for whether or not curvature is present and weighing it against the prior for larger flux scale uncertainties in the Haslam map also presents difficulty. The flux scale uncertainties are essentially a purely statistical quantity i.e. it is often sheerly expressed as a measure of how well the temperature data compares to whatever is considered a trustworthy reference, though sometimes it is physically defined such as in an absolutely calibrated survey. Whether spectral curvature is or is not present directly relates to the physics of the interstellar medium at these latitudes, and perhaps whether a point source with a nearby turnover has leaked into our smoothed data. One can of course make a measurement of these situations and subsequently construct a statistical model of the sky -- we are not saying the probability of curvature is immune to statistical reasoning. However, presenting these marginal likelihoods with a flat prior in a Bayesian setting reflects states of knowledge that we do not think are held by the community. 

In particular, a flat prior could reflect the state of knowledge of an agent who is totally ignorant about the presence of spectral curvature (sometimes referred to as the principle of indifference). It may also reflect the state of knowledge of an agent with positive knowledge of a very peculiar symmetry i.e. one that is convinced that exactly half the sky demonstrates spectral curvature. While the Bayesian evidence makes it clear that these data by themselves are not particularly in support of spectral curvature, a more informed participant might know that it is so overwhelmingly likely to be present\footnote{Perhaps strictly due to physical reasoning alone, in which case, how should we quantify that as a probability? Which propositions do we compute the relative truth values of and compound into the appropriate statement?} that this might be considered weak. To summarize, we can show in Figure \ref{fig:hyp_post} how these data ought to sway these hypotheses one way or the other relative to some state of knowledge, but without significantly more context than what is at our disposal we hesitate provide an authoritative judgement of any of these hypotheses (even in the limited context of this portion of the sky). 

Ideally this analysis would be robust against the choice of reference field. While we see that both choices of reference produce the same choice of dominant hypothesis, the preferences among the lower-likelihood models can be rather different. For instance, using reference field II seems to disfavor the model with steep spectral indices, no curvature, and nominal flux scale uncertainties (the left-most, null, hypothesis) with an evidence ratio in the tens of thousands, whereas this discrepancy is only in the hundreds when reference field I is used. As another example, $\mathbf{d}_\text{II}$ seems unable to discriminate between $\mathcal{H}_\alpha$ and $\mathcal{H}_{\alpha c h}$, whereas $\mathbf{d}_\text{I}$ seems to express a mild preference in favor of $\mathcal{H}_\alpha$. Since we have no robust way of setting priors on these different hypotheses, we will explore the ramifications of some choices of the model below even if they have exceedingly low Bayesian evidence with respect to these data.

\section{Fits and Parameter Constraints}
\label{sec:fits}

A fully Bayesian estimate of the parameters is formed by marginalizing over the hypotheses. Denoting the full set of parameters as $\boldsymbol{\theta}$,
\begin{equation}
    P(\boldsymbol{\theta} | \mathbf{d}) = \sum_\mathcal{H} P(\boldsymbol{\theta}, \mathcal{H} | \mathbf{d}) = \sum_\mathcal{H} P(\boldsymbol{\theta}| \mathcal{H}, \mathbf{d})P(\mathcal{H} | \mathbf{d})
    \label{eq:full_bayes}
\end{equation}
where $P(\mathcal{H} | \mathbf{d})$ is plotted in Figure \ref{fig:hyp_post} with a flat prior over hypotheses. The full posterior over the parameters is therefore a mixture over the various hypotheses. Rather than assume a particular prior and show the full inference according to Equation \ref{eq:full_bayes}, we instead explore the different hypotheses and note the correspondence between the marginal likelihood and e.g. the fit quality. 

\subsection{Choice of Haslam map flux scale factor prior}
\label{sec:wide_haslam}

\begin{figure*}
    \centering
    \includegraphics[width=\linewidth]{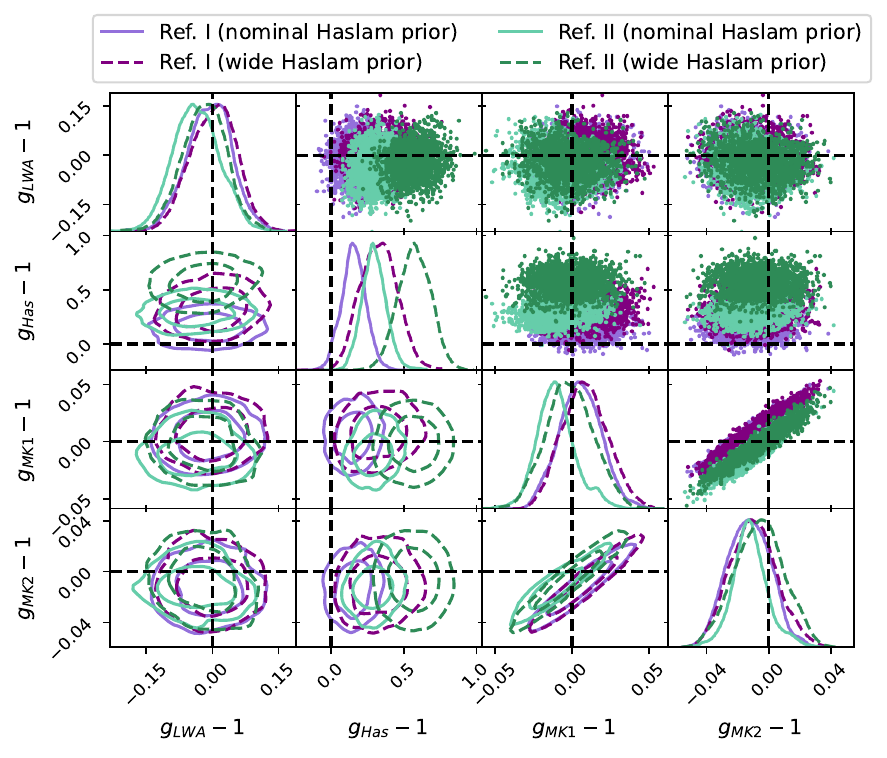}
    \caption{Inferred flux scale factors (offset by 1) depending on choice of reference field (color) and choice of Haslam map flux scale factor prior (linestyle). All models assume the narrow spectral index prior and no curvature. Positive values imply overestimates of the surface brightness of corresponding maps and vice versa for negative values. }
    \label{fig:gain_corner}
\end{figure*}

\begin{figure*}
    \centering
    \includegraphics[width=\linewidth]{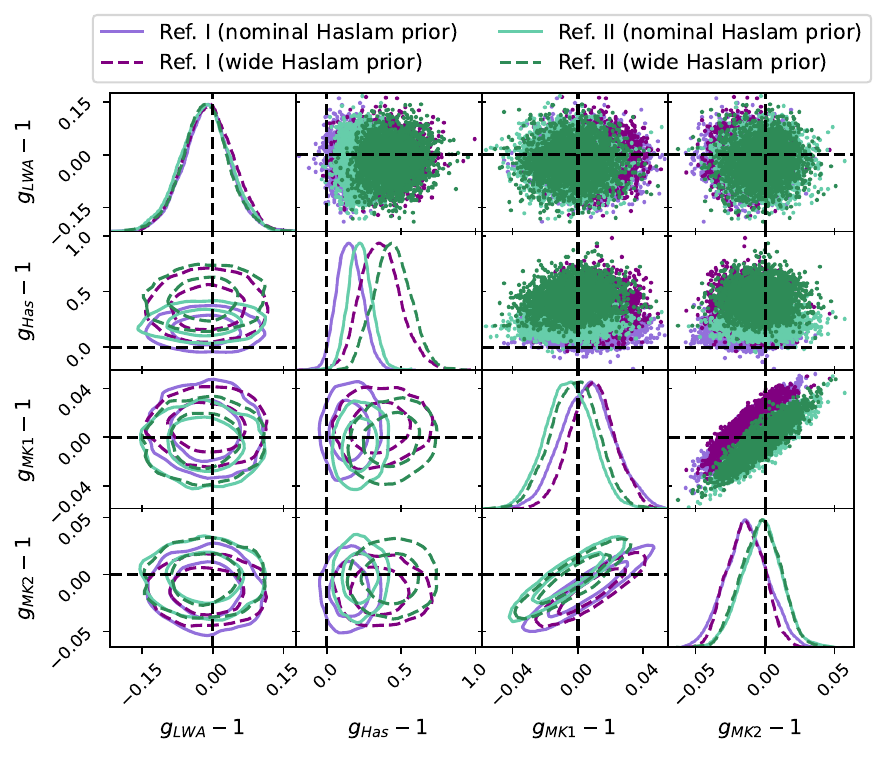}
    \caption{Same as Figure \ref{fig:gain_corner}, but for models that assume nonzero curvature.}
    \label{fig:gain_corner_curv}
\end{figure*}

We show a corner plot of the inferred flux scale factors depending on choice of reference field and Haslam flux scale factor prior in Figure \ref{fig:gain_corner}, with a narrow spectral index prior and no curvature. We have subtracted 1 from the flux scale factors to show fractional overestimations (positive values) or underestimations (negative values) i.e. flux scale factor offsets from the  corresponding experiments. The same style of plot is shown in Figure \ref{fig:gain_corner_curv}, but with curvature allowed; inferences are similar.

The posteriors for the OVRO-LWA and MeerKAT flux scale factors look very similar to their priors, though in some cases appear systematically offset from 1 by approximately a posterior standard deviation or two. The strength of the offset seems to depend more on choice of reference field than choice of Haslam map flux scale uncertainties. Since there is significant overlap of the marginal posterior of these flux scale factors with 1 (the flux scale factor offsets have significant overlap with 0), we do not assess these as particularly significant flux scale errors for these two experiments, though we do note mild evidence of discontinuity in the MeerKAT band since the two subband flux scale factors are slightly offset from one another. This is a known MeerKAT bandpass artifact with a standard correction strategy that was not implemented by \citet{Irfan2022} or us, since the correction strategy enforces certain assumptions about synchrotron emission that we explicitly wanted to test. In particular, it assumes a certain level of consistency with the Haslam map based on the Python Sky Model \citep{pysm}.   

On the other hand, the inferred Haslam map flux scale factor is significantly offset from 1 by at least two posterior standard deviations in all models considered in Figures \ref{fig:gain_corner} and \ref{fig:gain_corner_curv}. The positive offsets, which indicate an overestimation of the surface brightness, range from $15\% \pm 8\%$ to $57\% \pm 11\%$ depending on the hypothesis and choice of reference field. Models with spectral curvature tend to align with smaller inferred flux scale factors, since the negative spectral curvature can be used to make up for some of what would be an overestimated surface brightness in a straight power law model. Inferred flux scale factors between different choices of reference field present a systematic $\sim 1\sigma$ discrepancy, where reference field II seems to systematically produce higher inferred Haslam map flux scale factors regardless of the underlying hypothesis.  

When we adopt a wider prior on the Haslam map flux scale factor, the inferred flux scale factor tends to shift to more positive values. This suggests that the likelihood is concentrated at a relatively high value. We have not examined the likelihood directly, but a back of the envelope calculation assuming the likelihood has roughly Gaussian shape suggests that the maximum likelihood point for reference field II may be as high as $g_\text{Has} - 1 = 0.9$, which would suggest the Haslam map has overestimated the surface brightness in this part of the sky by a factor of two in the absence of regulation from a prior. The likelihood for reference I is probably concentrated at lower values, but still significantly offset from 1. Since the evidence is the average of the likelihood over the prior, likelihoods that are concentrated further from the center of the prior will tend to produce larger marginal likelihoods when larger \textit{a priori} uncertainties are assumed, until the increase in prior volume is so substantial that the tails of the likelihood are significantly contributing to the average. Examining hypotheses $\mathcal{H}_\alpha$ and $\mathcal{H}_{\alpha h}$ in Figure \ref{fig:hyp_post}, we can see this preference exemplified in the marginal likelihoods: use of reference field II expresses a stronger preference for wider flux scale factor priors than use of reference field I.

The range of inferred values and its sensitivity to choice of prior seems to demand some attention. We have only examined a small portion of the sky, so a more expansive study would provide important information on the following situations. A 15\% overestimation, when the reported uncertainty was 10\%, would not be particularly surprising. If this were the case, any analysis marginalizing over a flux scale factor with 10\% uncertainty would probably capture the structure of this nuisance in their astrophysical and/or cosmological parameter inferences. On the other hand, a 60\% (or 90\%) error is so large that it is likely to be underfit when a 10\% uncertainty is assumed unless there are sufficient data driving the fit such that discrepancies with the Haslam map can be ignored. 

\begin{figure*}
    \centering
    \includegraphics[width=\linewidth]{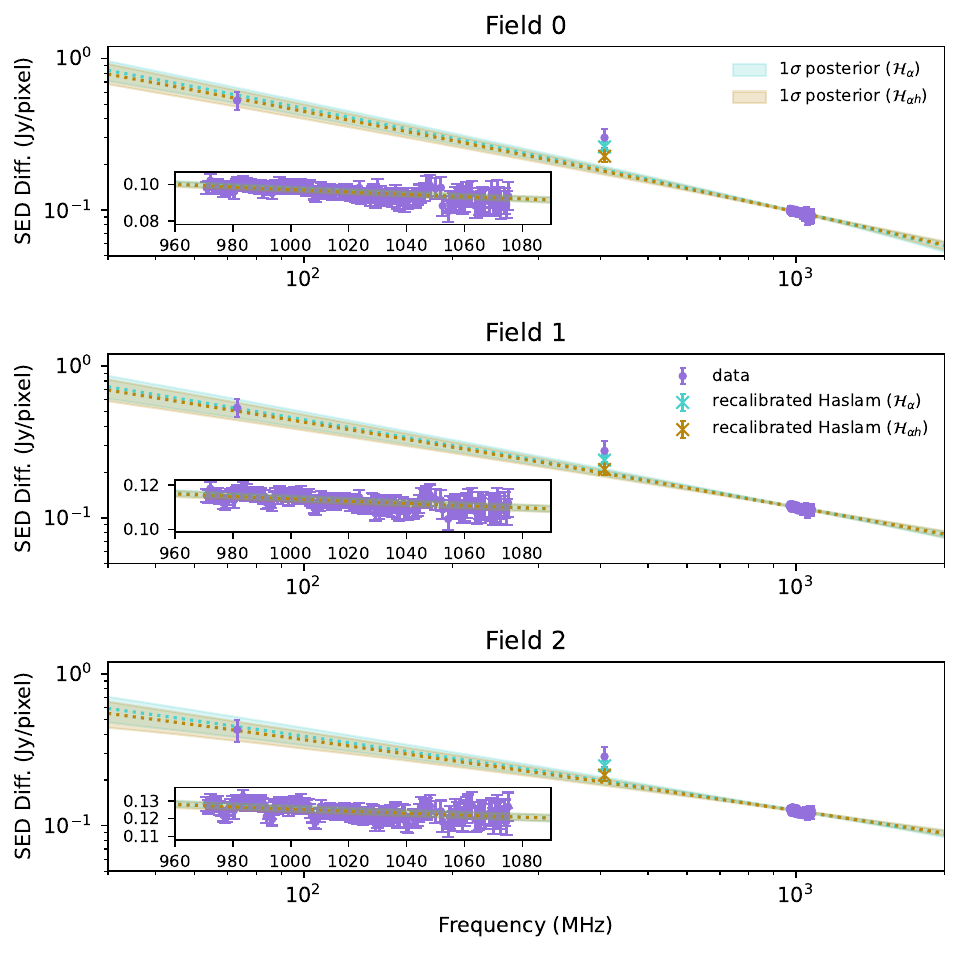}
    \caption{Inferred SED differences (Equation \ref{eq:sed_diff}) based on the inferred SED amplitude parameters ($S_i(\nu_0)$, $\alpha_i(\nu_0)$; spectral curvature is exactly 0 in these hypotheses) when reference field I is chosen for the analysis, whose spectral parameters were jointly inferred along with the analysis fields' parameters. We also show different choices of Haslam map flux scale factor priors, along with a rescaled Haslam points based on inferred flux scale factors.}
    \label{fig:SED_ref3}
\end{figure*}

\begin{figure*}
    \centering
    \includegraphics[width=\linewidth]{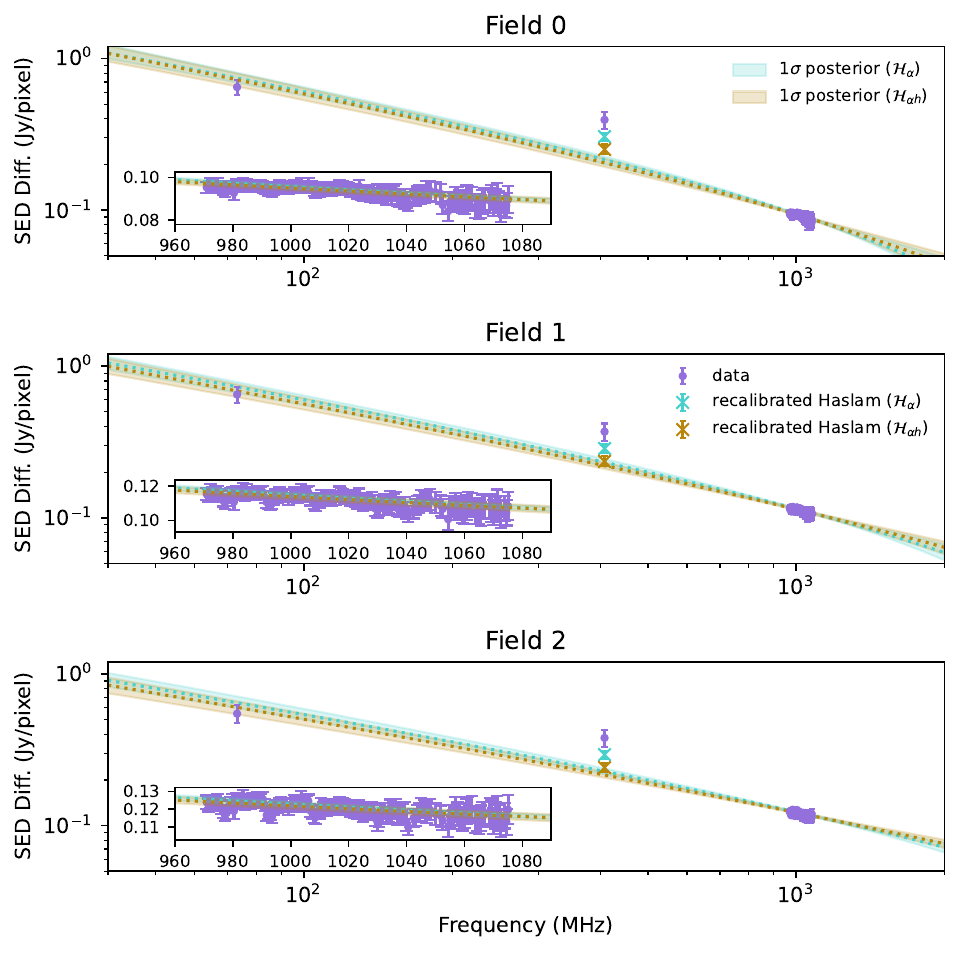}
    \caption{Same as \ref{fig:SED_ref3} but for when reference field II is chosen.}
    \label{fig:SED_ref4}
\end{figure*}

We show SED differences (Equation \ref{eq:sed_diff}) for each analysis field in Figures \ref{fig:SED_ref3} and \ref{fig:SED_ref4}, made separately by choosing reference fields I and II, respectively. We show two hypotheses in each panel of each figure; both adopt a narrow spectral index prior, but reflect different choices of Haslam map flux scale uncertainty (hypotheses $\mathcal{H}_\alpha$ and $\mathcal{H}_{\alpha h}$). In this case, regardless of the choice of flux scale factor prior, the SED difference seems to anchor to the OVRO-LWA and MeerKAT data. If we recalibrate the Haslam point by the inferred flux scale factor, we see that it better fits the SED differences, and that hypotheses with higher marginal likelihood produce better fits. This is generic regardless of the primary and reference apertures. The fit quality is even better if we allow for spectral curvature (not shown), but this increases the dimensionality of the model by four parameters (one curvature per analysis aperture plus one for the reference), ultimately leading to lower marginal likelihood. We suspect the spectra are largely insensitive to the Haslam data here because of the larger \textit{a priori} uncertainties and the fact that the Haslam data are at the center of the considered frequency range. Following this line of thinking, we wonder if inferred flux scale errors are biased towards being in the spectral center of a data set, but do not answer that question at this time. 

The importance of this result is that the Haslam map is commonly used as a low-frequency anchor in various aspects of cosmology. For example, it is used as a low-frequency model for Galactic synchrotron spectra in theoretical studies of foreground subtraction techniques for 21-cm cosmology \citep[e.g.][]{Shaw2014, Wolz2014, Bigot-Sazy2015, Olivari2016, Cunnington2019}. It is possible that the conclusions of such studies may be sensitive to large uncertainties in the overall amplitude of the map. Furthermore, if flux scale errors are spatially dependent, then forecasts may also be sensitive to the specific morphology of these errors. Observational cosmology results such as \citet{planck15x} that use Haslam as a low-frequency anchor may also be affected. In particular, foreground component separation, which is in part driven by the Haslam map, may obtain different results if different assumptions are made about the Haslam map's flux scale uncertainty. This would then have a knock-on effect for properties of the cosmological signal.

\subsection{Choice of synchrotron spectral behavior}

\begin{figure*}
    \centering
    \includegraphics[width=\linewidth]{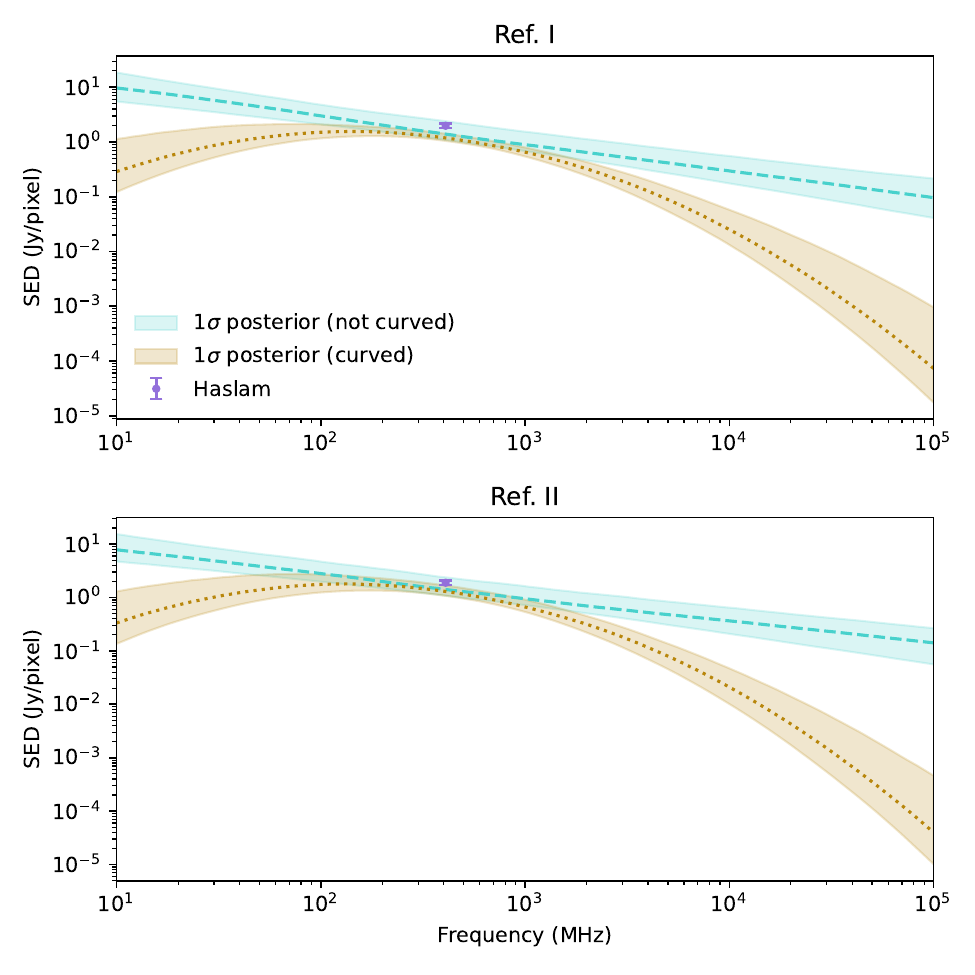}
    \caption{Inferred SEDs in the reference fields, ignoring large scale contributions, for the narrow spectral index prior when we do and do not allow for spectral curvature. We plot alongside the Haslam aperture data to demonstrate the large uncertainty in the amplitude parameters. Since neither the OVRO-LWA nor the MeerKLASS map have known zero levels, we do not plot them.}
    \label{fig:curv_compare}
\end{figure*}

Finally, we compare inferred SEDs (not SED differences, and neglecting large scale contributions; cf. Equation \ref{eq:SED}) in the reference fields when we do and do not allow spectral curvature (specifically hypotheses $\mathcal{H}_\alpha$ and $\mathcal{H}_{\alpha c}$). These are shown in Figure \ref{fig:curv_compare}. Inferred $\alpha_i(\nu_0)$ and $c_i$ seem insensitive to the width of the $\alpha_i(\nu_0)$ prior, so we omit that comparison. We therefore suspect the difference in marginal likelihood displayed in Figure \ref{fig:hyp_post} to be strictly due to an unnecessary increase in prior volume. When comparing the choice of flux scale factor prior among straight power law models, we naturally see a difference in the e.g. median prediction at any one frequency on the order of the size of the inferred Haslam flux scale factor, but the uncertainty in the $S_i(\nu_0)$ parameters essentially washes this out. We could perhaps make more precise statements about the SEDs using the alternate parametrization mentioned in \S\ref{sec:methods}.

The inferred curvature values in the reference fields for the synchrotron power laws depicted in Figure \ref{fig:curv_compare} are $-0.22 \pm 0.06$ and $-0.23 \pm 0.06$ for reference fields I and II, respectively. These posteriors (not shown) are mounded against the edge of the prior at $-0.3$, suggesting that if we extended the prior boundary, the estimates would support more negative values. This range of prior was chosen as a generous range compared to our understanding of Galactic synchrotron spectral curvature, but the appearance of the posterior suggests the data might support significant values outside of it. With this choice of prior boundary, the $2\sigma$ upper limits on the curvature are $-0.05$ (reference field I) and $-0.06$ (reference field II). Curvature estimates are similar in the analysis fields. The mean curvature estimates are, as far as we are aware, somewhat steeper than what is typically expected due to absorption processes in the interstellar medium. For example, Figure 15 of \citet{planck15xxv} shows that \texttt{galprop} \citep{Orlando2013} predicts a curvature of roughly -0.15 at these latitudes. Meanwhile, in the Galactic plane, \citet{kogut12} found a spectral curvature of just $-0.081 \pm 0.028$. Our posterior is inconsistent with these at $1\sigma$, however the curvature parameters are not particularly well-constrained, and these values are covered within the $2\sigma$ upper limits. 

A possible explanation for steep curvature is extragalactic background sources, since the maps were not desourced. Extragalactic sources with turnover appear somewhat more complicated than a curved power law, so the turnovers displayed in Figure \ref{fig:curv_compare} should not be taken literally in this interpretation. On the other hand, inferred curvatures are similarly steep in every single one of the five apertures, primary and reference alike. It seems unlikely that there are bright sources with spectral turnover near 100 MHz within all of the apertures. Instead we suspect that the model is just using curvature to drag the inferred Haslam flux scale factors back into a value of higher prior probability.  

\subsection{Potential cause of flux scale error in the Haslam map}

Where might such a large flux scale error arise in the Haslam map? With limited data, we can only speculate. An analysis conducted with several degree-scale absolutely calibrated maps of large sections of the sky, rather than just a few square degrees (and preferably with known zero levels to avoid this differencing strategy), could prove extremely useful in this regard.

In any case, we mention one plausible calibration error that could give rise to an effect of this size. The conversion from brightness temperature to antenna temperature is scale-dependent, as discussed by \citet{Jonas1998}. In particular, the solid angle of the source of interest must be taken into account when converting to brightness temperature. To account for this when studying diffuse emission, the data are often placed on the ``full-beam'' scale, meaning the antenna temperature must be scaled by the ratio of the ``main beam'' solid angle to the ``full beam'' solid angle. In \citet{Jonas1998}, a temperature conversion between two different beam scales is carried out via the formula 
\begin{equation}
    T_\text{FB} = T_\text{OFF} + \frac{\Omega_\text{A}}{\Omega_\text{FB}}T_\text{A}
\end{equation}
where \citet{Jonas1998} refer to $\Omega_\text{A}$ and $\Omega_\text{FB}$ as the ``total beam solid angle'' and ``full beam solid angle'', respectively. $T_\text{OFF}$ accounts for an uncalibrated zero level specific to that work. The main factor to consider here is the conversion $\Omega_\text{A}/\Omega_\text{FB}$, which is generally greater than 1. The total beam solid angle is defined in terms of the point-source sensitivity,
\begin{equation}
    \Omega_\text{A} = \frac{\lambda^2 \times \text{pss}}{2 k_B} \times 10^{-26}
\end{equation}
where $\lambda$ is the wavelength of the source, pss is the point-source sensitivity in Jy/K, and $k_B$ is Boltzmann's constant. The full beam scale, $\Omega_\text{FB}$, is constructed by integrating the beam over some solid angle. It usually does not refer literally to an integral out to the horizon, but rather out to some cutoff such as a few main beam widths \citep{Haslam1974, Reich1982, Jonas1998}. Meanwhile, one could convert to the main beam scale by integrating the beam out to the first null, providing $\Omega_\text{MB}$. Converting between the main beam and full beam scale would involve the ratio $\Omega_\text{MB} / \Omega_\text{FB}$. If the sidelobes are improperly characterized, or if the cutoff is chosen too close to boresight, then $\Omega_\text{FB}$ may be underestimated, resulting in an overestimated $T_\text{FB}$. If the map were not put on the full beam scale, or if the full beam scale were significantly underestimated, then one might observe a highly overestimated surface brightness consistent with our flux scale factor inference.

\citet{Haslam1974} claim only a 5\% difference in the two scales for the telescope used in that work (i.e. 95\% of the beam power is in the main beam). This claim is based on 10$^\circ$ sweeps surrounding Cas A (5$^\circ$ in either direction). Compared to other instruments, where a 30\% difference is more typical, this is a very low conversion factor that usually requires careful design and underillumination to acquire (see \S4.4 of \citet{Remazeilles:2014mba} and citations therein, as well as \citet{Dickinson2019}). It therefore seems unlikely that this 5\% figure is accurate, which makes this a plausible explanation for why we might see a flux scale error in the Haslam map as large as we do.

\subsection{Comparison to other results}

These results uniformly favor the hypothesis that the Haslam map is too bright to be consistent with the other data, and that there is unlikely to be any spectral curvature in these fields once this is corrected for. This result is seemingly inconsistent with \citet{Monsalve2021}, where globally scaling the Haslam map \textit{up} by 21\% and subtracting off a small 4.1 Kelvin additive offset improved agreement with absolutely calibrated EDGES data. Furthermore, allowing for spectral curvature had a similar effect on the Haslam map in that analysis (i.e. it produced agreement even without flux scale correction), whereas in our analysis we observe less degeneracy between these choices. We argue that these analyses cannot be directly compared in a simple way due to position-dependent complications in the Haslam map and a general incongruity between the analysis by \citet{Monsalve2021} and ours, although this discrepancy does need to be resolved to understand the nature of any correction that would need to be applied to the Haslam map.

The Haslam map is constructed from four different surveys made with different telescopes throughout the 60s and 70s \citep{Haslam82, Remazeilles:2014mba}. This makes it seem unlikely that a single rescaling over the entire sky should fix flux scale errors that may be arising from different physical sources specific to particular surveys. In fact, the discussion in section 5b of \citet{Haslam1981} claims different flux scale uncertainties for the Parkes survey data and Jodrell Bank survey data. In the simple case where there is a single rescaling procedure required for each independent survey, regions of overlap will require different rescalings to regions of non-overlap. Since we are analyzing a tiny region of the sky compared to the more global analysis of \citet{Monsalve2021}, it is unsurprising that these results may end up different. Moreover, given the nature of the Haslam map, it may be desirable to have a way of surgically searching for systematic flux scale errors in a position-dependent way, such as with \texttt{Chiborg}. This will be particularly relevant for foreground modeling methods that partition the sky such as that proposed by \citet{Pagano2024}.

In some fields on the sky, the synchrotron component can contribute a meaningful amount of brightness at the $\sim 30$ GHz spinning dust peak \citep{Dickinson2019, Rennie2022, Harper2024}. While our analysis is generally too uncertain to predict large, statistically significant distinctions at 30 GHz, in theory a 60\% flux scale error should overestimate contribution of Galactic synchrotron radiation to the spinning dust peak by the same amount. 

\subsection{Wide flux scale factor priors for all experiments}
\label{sec:wide_gain}

In this subsection, we consider a hypothesis where all experiments have underestimated their flux scale uncertainties (corresponding to overly narrow priors on the flux scale factors), there is no spectral curvature in the underlying synchrotron emission, and a narrow spectral index prior. Denoting this hypothesis as $\mathcal{H}_{\alpha g}$, we assume
\begin{equation}
    \mathbf{g} | \mathcal{H}_{\alpha g} \sim \mathcal{N}\left(\mathbf{1}_4, 0.2^2\mathbf{I}_4\right)
\end{equation}
i.e. that each experiment has a 20\% flux scale uncertainty. This model has nearly identical marginal likelihood as $\mathcal{H}_{\alpha h}$ for both choices of reference. This means both inferences are of equal weight in the absence of prior information discriminating between these hypotheses. We note that a flat prior in this instance indicates an equal belief that three experiments underestimated their flux scale uncertainties. If we imagine flux scale uncertainty mischaracterization as arising due to systematic effects within the experiment but whose natures vary from experiment-to-experiment, then it seems implausible for a large number of experiments for all of them to have underestimated them just by chance. While we only have a small number of experiments in this analysis, this line of thinking suggests there may be basic reasons to at least slightly disfavor $\mathcal{H}_{\alpha g}$ compared to $\mathcal{H}_{\alpha h}$. Furthermore, the calibration procedures of the respective maps are discussed in significantly more detail in \citet{Eastwood2018} and \citet{Wang2021}, compared to \citet{Haslam1981}, meaning at least subjectively it is harder to be skeptical of the uncertainty estimates provided by the LWA and MeerKAT experiments than the Haslam map. Nevertheless, the nearly equal marginal likelihoods are compelling, and the different priors implied by the different hypotheses produce markedly different inferences, which we deem important to explore. 

\begin{figure*}
    \centering
    \includegraphics[width=\linewidth]{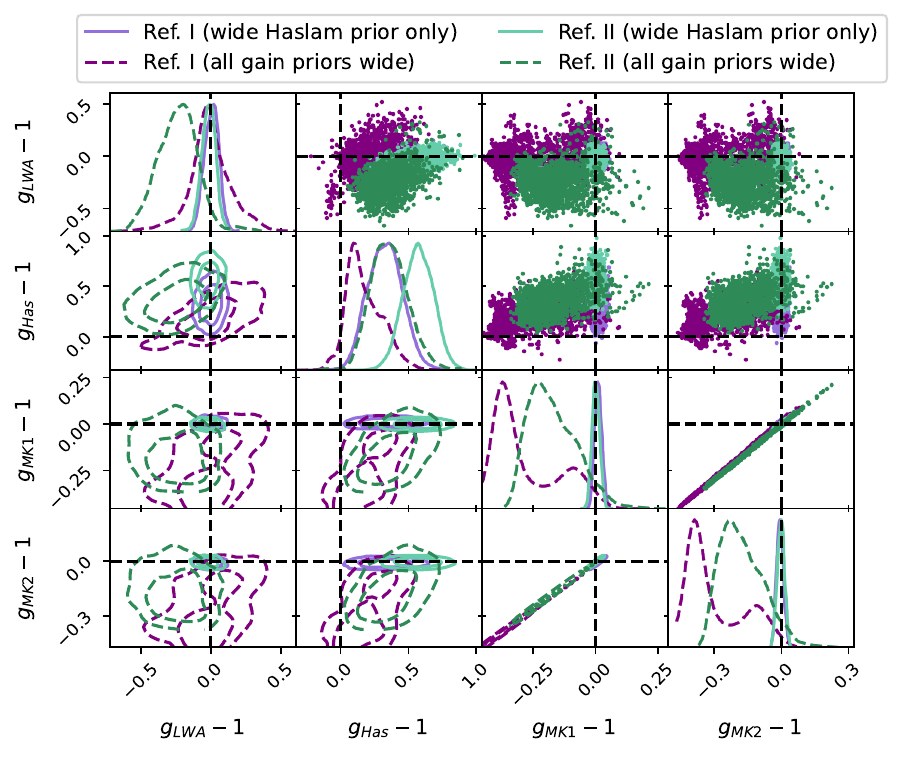}
    \caption{Flux scale factor posteriors for $\mathcal{H}_{\alpha h}$ (solid) and $\mathcal{H}_{\alpha g}$ (dashed), and different choice of reference. Assuming large flux scale uncertainties in all experiments removes the ability of any one experiment to ``anchor'' the others, resulting in multimodal inferences.}
    \label{fig:gain_all_wide}
\end{figure*}

We show a corner plot of the flux scale factor posteriors in Figure \ref{fig:gain_all_wide}. The posterior distributions reflect significantly higher uncertainty as a result of the more relaxed priors. The wide priors prevent any one experiment from serving as a reliable \textit{a priori} reference, which produces a complicated multimodal posterior, with more pronounced degeneracies between flux scale factors than in $\mathcal{H}_{\alpha h}$ (also shown in the figure). 

The multimodality is more obvious when reference field I is used, generating what appears to be two major modes and a potential minor one. The strongest mode suggests that the LWA and Haslam maps have flux scale factors roughly within the prior uncertainty, with a large $\sim$ 35\% underestimation in the MeerKAT data. The second strongest mode suggests a much larger positive flux scale factor offset in the Haslam map combined with a much smaller underestimation in the MeerKAT data. There appears to be a third, very small mode where the Haslam map has slightly underestimated the surface brightness.

The multimodality is less obvious when reference field II is used. It is only particularly visible as a shoulders in the marginal MeerKAT flux scale factor posteriors, which are hard to tell apart from sample errors in the kernel density estimation. As with $\mathcal{H}_{\alpha h}$, reference field II supports larger Haslam map flux scale factor posteriors, with little support for underestimations. The LWA flux scale factor posterior has a notably negative offset compared to reference field I. Due to the large uncertainties, the choice of reference field produces a only a mild tension as with all examined hypotheses, though the strong difference in posterior shape might as well be classified as a different type of tension. 

Weighting the inferences based strictly on the marginal likelihoods of the models would suggest that we have little understanding of where the flux scale errors exist in the full parameter space. However, it is crucial to note that the marginal likelihood only speaks to the degree to which this data supports any given hypothesis. As demonstrated by the toy example in \S\ref{sec:hyp_comp}, the strength of the prior can alter posterior probability of the model by orders of magnitude depending on the actual information defining the prior. The sensitivity of the analysis on the flux scale factor priors highlights the importance of well-documented flux scale uncertainty estimation. In other words, some of our confusion is a result of not knowing how strongly to believe the reported flux scale uncertainty estimates. Do we have a reason to believe the LWA flux scale uncertainty should be 20\%, rather than 5\%? We are significantly more skeptical of that possibility given what is presented in \citet{Eastwood2018} than we are of the corresponding possibility with the Haslam map. We recognize that scientific articles have become more detailed as time has progressed, and are not faulting \citet{Haslam1981} so much as pointing out that the progression towards increasing detail has been useful in future scientific work. This work suggests it may even be critical for combining measurements from different experiments.

\section{Conclusions}
\label{sec:conc}

Using a joint Bayesian analysis of data from three radio experiments (OVRO-LWA, the Haslam map, and MeerKLASS), we found significant evidence for a flux scale error in the Haslam map as high as 60\%. The presence and strength of this inferred flux scale error appears to depend on choices of reference field in our aperture photometry step. They also depend on whether we define the width of the flux scale factor priors via the reported flux scale uncertainties from each experiment versus adopting some arbitrary larger width to express more \textit{a priori} uncertainty. When assumed flux scale uncertainties are enlarged relative to experimental reports, we find some evidence for uncorrected flux scale factors in the LWA and MeerKAT data as well as the Haslam map. However, adopting broader flux scale factor priors for all experiments leads to degenerate inferences resulting in large \textit{a posteriori} uncertainties. The sensitivity of the analysis to different priors emphasizes the importance of understanding the \textit{a priori} strength of these competing models. This is to say, we can see how a flat prior over models in this problem may be unreasonable, and in that case, we need a way of setting the prior probabilities of the models in this context. 

We also compared whether any spectral curvature might be present. We found that when curvature is allowed, somewhat steep curvatures are preferred by the data, but only with modest certainty: $c \lesssim -0.15$ $(1\sigma)$, $c \lesssim -0.05$ $(2\sigma)$. We note that the curvature posteriors are mounded against the lower edge of the prior at $c=-0.3$, and these upper limits would likely be substantially reduced if we extended the range of the prior. In other words, the data likely support significantly steeper curvatures than these limits would suggest, and might be in tension with the current understanding of spectral curvature in Galactic synchrotron emission. However, the marginal likelihood of models involving curvature is generally quite low compared to corresponding alternatives, such as a broader \textit{a priori} distribution of flux scale factors in the Haslam map without any curvature. Curvature in this field is therefore disfavored statistically by these data, and also produces strange conclusions compared to physical expectations.

We emphasize that our model assumed that the only possible systematic effect were per-experiment flux scale errors (deviations of the flux scale of the map from the correct one by some multiplicative factor other than 1). However one could imagine different types of systematic effects masquerading as a large flux scale error in this model, e.g. an additive bias from strong radio frequency interference or an extragalactic point source. Indeed, the way that we difference the maps to perform the aperture photometry may be to blame if there exist variations on approximately $4^\circ$--$12^\circ$ scales that are unrelated to flux scale errors. This may cause a difference between the apertures across experiments that is inhomogeneous. In any case, calibration errors or other departures from the astrophysical model such as these can cause significant discrepancies in the inferred spectral properties of synchrotron emission. Errors in the spectral properties have direct relevance to inferences regarding the underlying astrophysical processes (e.g. the presence of curvature or lack thereof). For instance, the spectral curvatures we observe can change the contribution of synchrotron emission to the spinning dust peak at 30 GHz by more than an order of magnitude.

These discrepancies are also detrimental for cosmology experiments such as 21-cm intensity mapping surveys, which rely on foreground modeling for accurate calibration solutions. Even if an accurate calibration can be obtained through other means, such as \textit{in situ} calibration, an incorrect foreground model could cause structure in the foregrounds to be absorbed by parameters related to the cosmological signal unless appropriate precautions are taken, such as jointly modeling the foregrounds with sufficiently broad \textit{a priori} uncertainty \citep{Anstey2021, Pagano2024}. The possibility of such errors also makes spectral curvature and higher order spectral structure difficult to assess without a very broad range of frequencies. Applying this analysis or similar to a selection of compatible maps (e.g. with appropriate angular resolution) formed from spectroscopic measurements spanning a large range of frequencies would allow for the formulation of a radio sky model complete with formal flux scale uncertainties derived jointly among the contributing experiments. Such a sky model would help safeguard against inferential pitfalls in cosmology and astrophysics experiments that depend on accurate sky models. 

\section*{Acknowledgements}

We are grateful to the anonymous referee for their insightful requests. We acknowledge Clive Dickinson, Mohan Agrawal, and Robert Pascua for very helpful discussions. This result is part of a project that has received funding from the European Research Council (ERC) under the European Union's Horizon 2020 research and innovation programme (Grant agreement No. 948764; MW, PB). MW was funded by a CITA National Fellowship. MI acknowledges support from the South African Radio Astronomy Observatory and National Research Foundation (Grant No. 84156). PB acknowledges support from STFC Grant ST/X002624/1.

The MeerKAT telescope is operated by the South African Radio Astronomy Observatory, which is a facility of the National Research Foundation, an agency of the Department of Science and Innovation.

\section*{Data Availability}

The OVRO-LWA and (destriped) Haslam diffuse radio maps are available in the NASA-hosted LAMBDA archive.\footnote{\url{https://lambda.gsfc.nasa.gov/product/foreground/fg_diffuse.html}} Access to the raw MeerKAT data used in the analysis is public
(for access information please contact archive@ska.ac.za). Data products are available on request. The python code we wrote for our analysis is available at \url{https://github.com/mwilensky768/sed_jackknife}.



\bibliographystyle{mnras}
\bibliography{bayesiansed}

\bsp	
\label{lastpage}
\end{document}